\documentclass[aps, a4paper, superscriptaddress, nofootinbib, 10pt]{revtex4}
\usepackage{amsmath,amssymb, bm}
\usepackage{dsfont}
\usepackage{epsfig}
\usepackage{graphicx}
\usepackage{slashed}
\usepackage{color}
\usepackage{subfigure} 
\usepackage[colorlinks,citecolor=blue,urlcolor=blue,linkcolor=blue]{hyperref}

\newcommand{\be}{\begin{equation}}
\newcommand{\ee}{\end{equation}}

\newcommand{\beq}{\begin{equation}}
\newcommand{\eeq}{\end{equation}}
\newcommand{\bea}{\begin{eqnarray}}
\newcommand{\eea}{\end{eqnarray}}
\newcommand{\ba}{\begin{align}}
\newcommand{\ea}{\end{align}}
\newcommand{\bfig}{\begin{figure}}
\newcommand{\efig}{\end{figure}}

\begin{document}
~\vspace{1cm}
\title{Resurgent representation  of the Adler function in the large-$\beta_0$ approximation of QCD}
\author{Irinel Caprini}
\affiliation{Horia Hulubei National Institute for Physics and Nuclear Engineering, P.O.B. MG-6, 077125 Bucharest-Magurele, Romania}

\begin{abstract}  
Using a full resummation of the  Adler function  in the large-$\beta_0$ approximation of QCD   and a mathematical framework of resurgence suitable for the specific properties of the Borel transform in this particular case, we derive a compact  resurgent representation of the QCD Adler function, valid in the whole complex momentum plane. The representation is expressed  in terms of the inverse Mellin transform of the Borel function and is analytic in the complex  momentum plane, except for cuts along the timelike axis and the Landau region of the spacelike axis. It contains a purely nonperturbative term singular at the origin of the coupling plane, depending on a single real arbitrary constant. We compare the resurgent Adler function derived in this work  with a previous determination in a similar framework and use its values in the complex plane for a calculation of the hadronic width of the $\tau$ lepton in the Standard Model.
\end{abstract}

\maketitle
\vspace{0.5cm} 
\section{Introduction}
 Since the pioneering work of Dyson \cite{Dyson:1952tj}, it is widely accepted that perturbation theory  for the Green functions in quantum field theory leads to divergent series, which can be at most asymptotic to the expanded functions. This fact gives rise to nontrivial problems in the application of perturbation theory:  if the series were  convergent, the knowledge of all the perturbative coefficients  would uniquely determine the function. Uniqueness is lost however when dealing with divergent series: there are infinitely many functions having the same  asymptotic expansion.

A divergent power series indicates the fact that the expanded function is singular at the expansion point. This means that the Green functions of field theory are expected to be singular at the origin of the  coupling  plane. In the case of QED, the singular behavior was discovered by Dyson \cite{Dyson:1952tj}. For QCD, the existence of the singularity at zero coupling was demonstrated by 't Hooft \cite{tHooft}, using arguments based on unitarity, analyticity and renormalization-group invariance. The divergence can be inferred alternatively from particular classes of Feynman diagrams, which indicate a factorial growth of the expansion coefficients in both QED \cite{Lautrup:1977hs, Broadhurst:1992si} and QCD \cite{Beneke:1994qe, Beneke:1992ch}.  

The reconstruction of a
function from a factorially divergent power series is a difficult task. Borel summation is known to be a useful tool for dealing with divergent series. The large-order increase of the expansion coefficients of a function is encoded in the singularities of its Borel transform in the Borel plane. For some QCD correlators, like the Adler function, it is known that these singularities are placed along the negative axis of the Borel plane (ultraviolet renormalons) and along the positive axis (infrared renormalons and instantons) \cite{Beneke:1998ui}. Due to the singularities situated on the positive axis, in particular the infrared renormalons associated to the low-momenta contribution to the Feynman diagrams, the Laplace-Borel integral by which the function is recovered from the Borel transform is ambiguous. In mathematical language, the QCD Green functions are not Borel summable. 

In the mathematical theory of Borel non-summable series,  the sought function is expected to ``resurge`` by supplementing the perturbation series with terms singular at the expansion point, which cannot be seen perturbatively. These terms, denoted generically as ``nonperturbative'', can be organized sometimes as a ``transseries'', i.e. a sequence of power series, the expansion parameter of each of them being exponentially suppressed with respect to that of the previous series. Remarkably, insights on the form of the additional terms can be obtained from perturbation theory itself, more exactly from the large-order behavior of the perturbative asymptotic expansion.  

 There are several  different mathematical approaches to resurgence and transseries (see \cite{Ecalle1993, BerryHowls, Howls, Dorigoni:2014hea, Costin1995, Costin_Duke, Costin_Book,  Aniceto:2018bis, Reis:2022tni} and references therein).   Some methods start from the original power series truncated at the lowest term, others require the knowledge of the full perturbation series. Besides the pure mathematical studies, applications to various models of quantum fieid theory have been also considered  \cite{Argyres:2012ka, Dunne:2012ae, Clavier:2019sph, Borinsky:2020vae, Abbott:2020qnl, Fujimori:2021oqg, Borinsky:2022knn}. However, there are few applications of the formal mathematical techniques to specific quantities in QCD. One such application is developed in a series of papers \cite{Ayala:2019uaw, Ayala:2019hkn, Ayala:2020pxq}, where the truncated perturbation series of some QCD correlators have been supplemented by nonperturbative terms related to  the first infrared renormalon, in the spirit of the hyperasymptotic formalism from \cite{BerryHowls, Howls}. Other application is the Borel-\'Ecalle resummation of the Adler function presented in  \cite{Maiezza:2021mry, Maiezza:2021bed}, based on the previous works \cite{Maiezza:2019dht, Bersini:2019axn}, where is was shown how the renormalization-group equation enables the applicability of the resurgence approach developed in \cite{Costin1995, Costin_Duke, Costin_Book}.  

 Although formal applications are rare,  attempts to complete the QCD perturbation series  by nonperturbative contributions  have been made  at a phenomenological level since a long time. The most popular example is provided by the power corrections present in the operator product expansions (OPE), introduced in the famous SVZ paper \cite{Shifman:1978bx} and exploited since then in a large number of works. 
 
 Taking into account the running of the strong coupling as a function of momentum squared, $\alpha_s(-q^2)\sim 1/\ln (-q^2)$, the powers of $1/(-q^2)$ can be written as  exponentially-supressed terms of the form $\exp(-1/\alpha_s(-q^2))$, and can be interpreted,  in mathematical language, as the first piece of a transseries  to the original perturbation expansion in powers of $\alpha_s$.  Going further, exponentially-decreasing corrections in the momentum plane have been assumed to exist (they are usually called ``quark-hadron duality-violating terms'' \cite{Shifman:2000jv}), and  can be viewed mathematically as yet another piece of the transseries. While in the mathematical approach the terms to be added to the perturbation series are predicted in a formal way, in the phenomenological studies of QCD a specific physical interpretation is  assigned usually to the additional terms. Thus, the power corrections are assumed to be due to the special properties of the QCD nonperturbative vacuum, expressed by nonzero quark and gluon condensates \cite{Shifman:1978bx}, while the duality-violating corrections appear to be required by the resonances in the hadron picture of the QCD correlators (see for instance \cite{Boito:2017cnp} and references therein). 

It is of interest to mention also the method of conformal mappings, which accelerates the convergence of power series by reorganizing them as expansions in powers of the variable that performs the  mapping of the analyticity domain of the expanded function onto a disk.  In QCD,  the method cannot be applied directly to the perturbative expansions in powers of the coupling, since the expanded functions are singular at the expansion point $\alpha_s=0$, but it can be used for the expansion of the corresponding Borel transforms, which are analytic in a region around the origin of the Borel plane. For the QCD Adler function, the optimal conformal mapping of the whole cut Borel plane onto a unit disk was investigated for the first time  in  \cite{Caprini:1998wg} and was applied  since then in many phenomenological studies (see for instance \cite{Caprini:2011ya, Caprini:2020lff}). They allowed in particular a more accurate extraction of the strong coupling from  hadronic $\tau$ decays and the  prediction of higher-order perturbative coefficients of the Adler function. For the present discussion, it is useful to emphasize that, as argued in \cite{Caprini:2020lff},  the method allows  to recapture some nonperturbative features of the exact function,  equivalent to some extent to the power corrections in OPE.

In the present paper, we shall discuss the application to QCD  of the mathematical resurgence approach developed in  \cite{Costin1995, Costin_Duke, Costin_Book}. As we mentioned above, the problem was considered recently in the literature: in \cite{Bersini:2019axn}, the authors argued that the renormalization-group equation obeyed by the QCD correlators allows one to apply the Borel-\'Ecale resummation  \cite{Costin1995, Costin_Duke, Costin_Book} for the resurgent solutions of nonlinear ordinary differential equations, and in \cite{Maiezza:2021mry} the method was applied to the particular case of the QCD Adler function. 
The mathematical framework requires  the knowledge of the whole perturbation series of the function of interest, more exactly the structure of the full Borel transform, whose singularities must be poles.  For the Adler function, there requirements are met in a certain limit of perturbative QCD, known as large-$\beta_0$ approximation \cite{Beneke:1992ch, Broadhurst:1992si, Beneke:1994qe}. Taking advantage of this fact, the authors of \cite{Maiezza:2021mry} obtained a resurgent representation of the Adler function which contains, besides the perturbation series, an additional term, calculated using the isomorphism presented in \cite{Costin_Duke}. In the subsequent study \cite{Maiezza:2021bed}, some additional assumptions have been adopted, aiming to make the resurgent representation useful for phenomenological applications

 In our work, we adopt the arguments presented in \cite{Bersini:2019axn, Maiezza:2019dht} on the applicability of the resurgent formalism  \cite{Costin1995, Costin_Duke, Costin_Book} to the QCD correlators, but bring several improvements and generalizations to the results presented in \cite{Maiezza:2021mry, Maiezza:2021bed} for the Adler function.  The crucial input which we shall use in our study is the compact resummation of the QCD Adler function in the large-$\beta_0$ limit, derived in \cite{Caprini:1999ma}. This resummation, apparently ignored by the authors of \cite{Maiezza:2021mry, Maiezza:2021bed}, leads to a representation of the Adler function in the complex momentum plane, expressed in terms of the inverse Mellin transform of the Borel function, which turns out to be related to the discontinuity of this function across the cuts in the Borel plane. On the other hand, the discontinuity across the positive axis is the main ingredient of the nonperturbative term calculated by the resurgent algorithm of \cite{Costin1995, Costin_Duke, Costin_Book}. This allows us to obtain a compact resurgent representation of the Adler function,  valid in the whole complex momentum plane, not only on the spacelike axis as in \cite{Maiezza:2021mry, Maiezza:2021bed}. Moreover, our result includes in a complete way the contribution of all renormalons, while in  \cite{Maiezza:2021mry}  the ultraviolet renormalons are omitted. Finally, we avoid some assumptions adopted in \cite{Maiezza:2021mry, Maiezza:2021bed} concerning the lowest order perturbative coefficients and the running of the strong coupling. We obtain therefore the exact solution of the resurgence problem for the Adler function in the large-$\beta_0$ limit of QCD.

The outline of the paper is as follows: in the next section we briefly review the compact resummation of the Adler function in QCD in the large-$\beta_0$ approximation, derived in \cite{Caprini:1999ma}. In Sec. \ref{sec:resur} we use this resummation and  the techniques developed in \cite{Costin1995, Costin_Duke, Costin_Book} to derive the exact resurgent representation of the Adler function, which depends on a single arbitrary real constant. In Sec. \ref{sec:tau} we illustrate the validity of this representation   in the complex momentum  plane  with a calculation of the total hadronic width of $\tau$ lepton. Finally,  Sec. \ref{sec:conc} contains a brief discussion and our conclusions.

\section{QCD Adler function in the large-$\beta_0$ approximation}\label{sec:resum}
We consider the QCD Adler function  $D(s)$  defined as
\begin{equation}\label{eq:Ddef}
   D(s) = -s\,\frac{d\Pi(s)}{ ds},\quad\quad s=q^2,
\end{equation}
in terms of the current--current correlation amplitude $\Pi(q^2)$ defined by
\begin{equation}\label{eq:2point}
   -4\pi^2 i\int d^4 x\,e^{i q\cdot x}\,\langle 0|\,T\,
   \{ J^\mu(x) (J^\nu(0))^\dagger \}\,|0\rangle
   = (q^\mu q^\nu - g^{\mu\nu} q^2)\,\Pi(q^2),  
\end{equation}
where $J^\mu$ is the vector current for massless quarks. 
  From the general principles of field theory, it is known that $D(s)$ is an analytic function of real type (i.e., it satisfies the  Schwarz reflection property $D(s^*)=D^*(s)$, where the star denotes complex conjugation) in the complex $s$ plane cut along the timelike axis for $s\ge 4 m_\pi^2$. 

 In renormalization-group-improved perturbation theory,  $D(s)$ is expressed formally as an expansion 
\beq\label{eq:pertD}
D(s) =1+\sum\limits_{n\ge 1} \left[\frac{\alpha_s(-s)}{\pi}\right]^n \,c_{n,1},
\eeq
in powers of the renormalized strong coupling $ \alpha_s(-s)$   in a certain renormalization scheme.
The  perturbative coefficients  $c_{n,1}$ in this expansion are obtained from Feynman diagrams (in full QCD calculations are available at present up to $\alpha_s^4$, see \cite{Baikov:2008jh} and references therein).
On the other hand, at high orders the perturbative coefficients  are known to increase factorially,   more exactly $c_{n,1}\approx K\,n^{a}\,b^n\, n !$,  where  $K$, $a$ and $b$ are  constants  \cite{Beneke:1998ui}.  Therefore, the series (\ref{eq:pertD})  has zero radius of convergence and can be interpreted only as an asymptotic expansion to $D(s)$ for $\alpha_s\to 0$. This indicates the fact that the Adler function, viewed as a function of the strong coupling $\alpha_s$, is singular at the origin $\alpha_s=0$ of the coupling plane. 

In some cases, the expanded function can be recovered  from its divergent power expansion through Borel summation. We define the Borel transform associated to the expansion (\ref{eq:pertD}) as
\be\label{eq:B}
 \widehat B_D(u)= \sum_{n=0}^\infty  b_n\, u^n,
\ee
where the coefficients $b_n$ are related to the perturbative coefficients $c_{n,1}$ by 
\be\label{eq:bn}
 b_n=\frac{4^{n+1}\,c_{n+1,1} }{\beta_0^n \,n!},\quad n\ge 0.
\ee
Then one can check that the function $D(s)$ is given formally by the Laplace-Borel integral
\begin{equation}\label{eq:Laplace}
   D(s) = 1 + \frac{1}{\beta_0} \int\limits_0^\infty\! du
    \exp\left( -\frac{4\pi u}{\beta_0\alpha_s(-s)} \right)\widehat B_D(u).
\end{equation}

 The meaning of this representation depends on the analytic properties of the expanded function in the complex $\alpha_s$ plane  or, alternatively, on the analytic properties of the Borel transform $\widehat B_D(u)$ in the $u$ plane.
In the mathematical theory of Borel summation, criteria are given for recovering an analytic function with desired properties from the Borel-Laplace representation. As we already mentioned, these criteria are not met in the case of QCD correlators like the Adler function (see for instance the review \cite{Fischer:1994mm}). 

We note first that, due to the factor $n!$ in the denominator of (\ref{eq:bn}), the series (\ref{eq:B}) in expected to have a nonzero radius of convergence. Therefore,
the singularities of the Borel transform are expected to be placed at a nonzero distance from the origin $u=0$ of the Borel plane. Actually, these singularities   encode the large-order increase of the coefficients of the original series (\ref{eq:pertD}). In the case of the Adler function, it is known that   $\widehat B_D(u)$ has singularities, assumed to be in general branch points, at integer values of $u$ on the semiaxes $u\ge 2$ (infrared renormalons and instantons) and $u\le -1$ (ultraviolet renormalons). The  nature of the first singularities, closest to the origin $u=0$, has been determined from renormalization-gruop arguments \cite{Mueller1985, Beneke:1997qd, Beneke:1998ui, Beneke:2008ad}, but  the corresponding residues and the nature and residues of the other singularities are not known.

We recall that several models of the Adler function have been proposed in the literature \cite{Beneke:2008ad, 
Beneke:2012vb, Cvetic:2018qxs}, which  use a description in terms of a small number of leading renormalon singularities of the Borel transform. Although these models proved to be useful for guiding the phenomenological applications or for testing the expansions based on conformal mappings \cite{Caprini:2011ya}, they are affected by uncertainties, especially due to the fact that the residues of the first renormalons are not known.  

In the present paper we will be interested in an approximation of QCD known as large-$\beta_0$ limit, obtained when only terms with the highest power in the number of light fermons are kept in the Feynman diagrams. Correspondingly, the running of $\alpha_s$  is determined by the one-loop renormalization-group equation (for details of the large-$\beta_0$  approximation see  \cite{Beneke:1992ch, Broadhurst:1992si, Beneke:1994qe, Beneke:1998ui}). In this case, the coefficients $c_{n,1}$ are known to all-orders and exhibit a factorial growth (see Table 1 of \cite{Beneke:2008ad}, which lists the coefficients up to 12th order). For convenience, we quote the first five coefficients in the 
$\overline{\text{MS}}$ renormalization scheme:
 \beq
 c_{1,1}=1,\quad   c_{2,1}=1.557, \quad c_{3,1}=15.711, \quad c_{4,1}=24.832,
 \eeq
 to be compared with the exact coefficients $1, \,1.64,\, 6.371, \,49.07$ calculated in the same scheme in full QCD,  cf. \cite{Baikov:2008jh} and references therein. 
 
 In the large-$\beta_0$ limit, the exact expression  of the Borel transform $\widehat B_D(u)$  is known \cite{Beneke:1992ch, Broadhurst:1992si}:
\begin{equation}\label{eq:BDu}
   \widehat B_D(u) = \frac{128\ e^{-C u}}{3(2-u)}\,\sum_{k=2}^\infty\,
   \frac{(-1)^k\,k}{\big[ k^2-(1-u)^2 \big]^2},
\end{equation}
where $C$ is a scheme-dependent constant, which compensates the scheme dependence of the coupling to ensure the scheme independence of 
the Laplace-Borel integral (\ref{eq:Laplace}). We recall that $C=-5/3$ in the $\overline{\text{MS}}$ renormalization scheme and $C=0$ in the so-called $V$ scheme \cite{Neubert:1994vb}, which we shall adopt in this work.  We use for consistency  the running strong coupling $\alpha_s(-s)$  to one-loop
\begin{equation}\label{eq:asrng}
   \alpha_s(-s) = \frac{4\pi}{\beta_0\ln(-s/\Lambda^2_V)},
\end{equation}
where the QCD scale parameter $\Lambda_V$  in the   $V$ scheme is related to  $\Lambda_{\overline{\text{MS}}}$ by \cite{Neubert:1994vb}
\beq\label{eq:Lambda}
\Lambda_V= e^{5/6}\,\Lambda_{\overline{\text{MS}}}.
\eeq
We recall that $\beta_0$ in (\ref{eq:asrng}) is  the first coefficient 
of the perturbative expansion of the $\beta$-function which governs the $\alpha_s$-running,  in the convention $\beta_0  =\frac{11}{3} N_c - \frac 23 n_f$, where $N_c=3$ and $n_f=3$. One notes in (\ref{eq:asrng}) the Landau pole at $s=-\Lambda_V^2$,  present also in the perturbative expansions (\ref{eq:pertD}) truncated at finite orders.

From  (\ref{eq:BDu}) it is seen that $\widehat B_D(u)$ has   singularities along the real axis, which are poles (unlike the case of full QCD, where the singularities are expected to be branch points). Due to the singularities on the positive axis $u>0$, the integral  (\ref{eq:Laplace}) is ill-defined and requires a regularization.  We consider the principal value (PV) prescription, defined as
\begin{equation}\label{eq:PV}
   D_{\text{PV}}(s) 
   =1 + \frac{1}{2\beta_0} \int_{\cal C_+}\! du
    \exp\left( -\frac{4\pi u}{\beta_0\alpha_s(-s)} \right)\widehat B_D(u) +\frac{1}{2\beta_0} \int_{\cal C_-}\! du
    \exp\left( -\frac{4\pi u}{\beta_0\alpha_s(-s)} \right)\widehat B_D(u),
\end{equation}
 where ${\cal C_\pm}$  are parallel lines 
slightly above and below the positive axis. This prescription is a generalization of the Cauchy principal
value for simple poles, and is the only choice giving a real
value for the Borel-summed amplitude when the coupling constant is real.  In \cite{Caprini:1999ma} it was argued that the definition (\ref{eq:PV}) is more consistent than other prescriptions with the analytic properties of the Adler function in the complex $s$ plane, which follow from causality and unitarity in field theory.  In particular, the Adler function defined by (\ref{eq:PV}) satisfies Schwarz reflection property, $D(s^*)=D^*(s)$, and is real along the spacelike axis $s<0$.  We shall adopt therefore in this work the PV prescription (\ref{eq:PV}), and for simplicity omit in what follows the subscript PV in the notation of the Adler function.  

A closed compact resummation of the Adler function in the large-$\beta_0$ limit was obtained in \cite{Caprini:1999ma}, using the technique of Mellin transform. The method is based on the representation 
\begin{equation}\label{eq:wninv}
   \widehat B_D(u) = \int\limits_0^\infty\!d\tau\,
   \widehat w_D(\tau)\,\tau^{-u}, \quad\quad u\in(-i\infty,\,i\infty),
\end{equation}
of the Borel transform, in terms of the inverse Mellin transform  $\widehat w_D(\tau)$, defined as
\begin{equation}\label{eq:wDn}
   \widehat w_D(\tau) = \frac{1}{2\pi i}
   \int\limits_{u_0-i\infty}^{u_0+i\infty}\!du\,
   \widehat B_D(u)\,\tau^{u-1}.
\end{equation} 

The function  $\widehat w_D(\tau)$   was introduced for the first time in \cite{Neubert:1994vb}, where its physical interpretation as  ''distribution function`` of the internal gluon virtualities
in Feynman diagrams was pointed out. It can be calculated from 
(\ref{eq:wDn}) by closing the integration contour along a semicircle at
infinity in the $u$ plane and applying the theorem of residues for the 
singularities of $\widehat B_D(u)$ located inside the integration 
domain. For $\tau<1$ the contribution from the 
semicircle at infinity vanishes if the contour is closed in the 
right half of the $u$ plane, while for $\tau>1$ the contour must be
closed in the left half plane. One thus obtains different 
expressions for the distribution function depending on whether
$\tau<1$ or $\tau>1$, which we shall denote, as in  \cite{Caprini:1999ma}, by $\widehat 
w_D^{(<)}(\tau)$ and $\widehat w_D^{(>)}(\tau)$, respectively. 
By Cauchy's theorem, these functions are then related to the discontinuity of the Borel transform across the real axis.  For instance,
$\widehat w_D^{(<)}(\tau)$ has the expression
\begin{equation}\label{eq:wn1}
   \widehat w_D^{(<)}(\tau) = \frac{1}{2\pi i}\int_{\cal C_+}\!
   du\,\widehat B_D(u)\,\tau^{u-1} - \frac{1}{2\pi i}
   \int_{\cal C_-}\!du\,\widehat B_D(u)\,\tau^{u-1} \,,
\end{equation}
where the contours ${\cal C_+}$ and ${\cal C_-}$ are lines slightly 
above and below the positive real axis of the $u$ plane. A similar 
representation is obtained for the function $\widehat w_D^{(>)}(\tau)$, where 
the integration lines are now parallel to the negative axis.
 In 
the large-$\beta_0$ limit, the evaluation of these representations requires the summation over the residues of all the infrared and unltraviolet renormalons of the Borel transform, which leads to the expressions \cite{Neubert:1994vb}
\begin{eqnarray}\label{eq:wDfun}
   \widehat w_D^{(<)}(\tau) &=& \frac{32}{3} \left\{ \tau\left(
    \frac74 - \ln\tau \right) + (1+\tau)\Big[ L_2(-\tau) + \ln\tau
    \ln(1+\tau) \Big] \right\}, \\
   \widehat w_D^{(>)}(\tau) &=& \frac{32}{3} \left\{ 1 + \ln\tau
    + \left( \frac34 + \frac12 \ln\tau \right) \frac{1}{\tau}
    + (1+\tau)\Big[ L_2(-\tau^{-1}) - \ln\tau \ln(1+\tau^{-1}) \Big]
    \right\}, \nonumber
\end{eqnarray}
where $L_2(x)=-\int_0^x \frac{d t}{t}\ln(1-t)$ is the dilogarithm.
As remarked in \cite{Caprini:1999ma}, the above expressions are analytic in the complex $\tau$ plane, with
no singularities other than branch cuts along the negative real
axis  $\tau<0$. Taken together, they define a function $\widehat w_D(\tau)$ that
is piece-wise analytic in the cut $\tau$ plane, with different 
functional expressions for $|\tau|<1$ and $|\tau|>1$. 

A compact form for the Adler function is obtained by introducing in (\ref{eq:PV}) the representation (\ref{eq:wninv}) of the Borel transform. Actually, since this representation is valid along the imaginary axis of the $u$-plane, one must first rotate the integration contour from the lines ${\cal C_\pm}$ up to the imaginary axis, which is possible if the circle quadrants at large $|u|$ bring a negligible contribution.  From (\ref{eq:PV}) one can see that the behavior of the integrand at large $|u|$ depends on the imaginary part of $u$ and the imaginary part of the variable $s$ entering the one-loop coupling (\ref{eq:asrng}).  A crucial remark is that, when the proper behavior is not achieved along a circle quadrant, one must first cross the positive axis of the $u$ plane and perform the rotation in the opposite quadrant, where the integral along the circle can be neglected. However, due to the singularities of the Borel transform, when crossing the positive axis one picks up an additional contribution, which can be calculated using (\ref{eq:wn1}).  The careful analysis of the various cases, performed in \cite{Caprini:1999ma}, leads  to
the following expression for the Adler function in  the complex $s$-plane:
\begin{equation}\label{eq:duplo}
   D^{(\pm)}(s) = 1 + \frac{1}{\beta_0} \int\limits_0^\infty\!
   d\tau\,\frac{\widehat w_D(\tau)}{\ln(-\tau s/\Lambda_V^2)} 
   \mp \frac{i\pi}{\beta_0} \left( -\frac{\Lambda_V^2}{s} \right)
   \widehat w_D^{(<)}(-\Lambda_V^2/s),
\end{equation}
where the superscript ``$\pm$'' in parenthesis refers to the sign of 
  $\text{Im}\,s$. 
  
As discussed in detail in \cite{Caprini:1999ma},  the functions $D^{(\pm)}(s)$ are 
holomorphic for complex values of $s$ in the 
upper and lower half planes, outside the real axis, where the integral in (\ref{eq:duplo}) is well defined. Moreover, it is easy to check that
\beq\label{eq:Schwarz}
 D^{(+)}(s^*) = (D^{(-)}(s))^*.
\eeq

 The values on the real axis are found by taking the limit of (\ref{eq:duplo}) for  $\text{Im}\,s \to 0$.  One obtains in particular the imaginary part of $D^{(+)}(s)$ on the upper edge of the timelike axis  \cite{Caprini:1999ma}
\begin{equation}\label{eq:imagdti}
   \mbox{Im}\,D^{(+)}(s+i\epsilon) = \frac{\pi}{\beta_0} \left[\,
   \int\limits_0^\infty\!d\tau\,
   \frac{\widehat w_D(\tau)}{\ln^2(\tau s/\Lambda_V^2) +\pi^2}
   + \frac{\Lambda_V^2}{s}\,\mbox{Re}\,\widehat w_D^{(<)}
   (-\Lambda_V^2/s) \right], \quad s>0,
\end{equation}
and on the upper edge of the spacelike axis  \cite{Caprini:1999ma}
\begin{equation}\label{eq:imagdsp}
   \mbox{Im}\,D^{(+)}(s+i\epsilon) = \frac{\pi}{\beta_0} \left(
   -\frac{\Lambda_V^2}{s} \right) \left[ \widehat w_D(-\Lambda_V^2/s)
   - \widehat w_D^{(<)}(-\Lambda_V^2/s) \right],\quad s<0.
\end{equation}
We recall that the functions $\widehat w_D^{(>)}$ and $\widehat w_D^{(<)}$ are real for positive arguments, therefore the two terms in (\ref{eq:imagdsp}) are real. 
If $s<-\Lambda_V^2$, the arguments of the 
functions $\widehat w_D(\tau)$ and $\widehat w_D^{(<)}(\tau)$ 
appearing in (\ref{eq:imagdsp}) are less than one. For such arguments, the 
function $\widehat w_D$ coincides with $\widehat w_D^{(<)}$ and the two terms in (\ref{eq:imagdsp}) 
compensate each other. Therefore, the imaginary part of the function $D^{(+)}(s)$ vanishes for spacelike momenta 
outside the Landau region:
\begin{equation}\label{eq:imagdsp1}
   \mbox{Im}\,D^{(+)}(s+i\epsilon) = 0,\quad s<-\Lambda_V^2.
\end{equation}
On the other hand, for $-\Lambda_V^2<s<0$, the arguments of the functions in (\ref{eq:imagdsp}) are greater than one, when $\widehat w_D$ coincides with $\widehat w_D^{(>)}$. Then the two terms in (\ref{eq:imagdsp}) do not compensate each other and give a nonzero imaginary part of $D^{(+)}(s+i\epsilon)$.

The same results are obtained for  $D^{(-)}(s)$  when approaching the real  axis 
from below. It follows that, for $s<-\Lambda_V^2$ both $D^{(+)}(s)$ and $D^{(-)}(s)$ are real and are given by
\begin{equation}\label{eq:deucl1}
   D^{(\pm)}(s) = 1 + \frac{1}{\beta_0}\,\mbox{Re} \int\limits_0^\infty\!
   d\tau\,\frac{\widehat w_D(\tau)}{\ln(-\tau s/\Lambda_V^2)},\quad s<-\Lambda_V^2.
\end{equation}
 One can use then Eqs. (\ref{eq:Schwarz})  and (\ref{eq:deucl1}) and the Schwarz reflection principle to show that  $D^{(\pm)}(s)$ define a unique function $D(s)$ real-analytic in the complex  $s$  plane.  This function is real on the spacelike axis for $s<-\Lambda_V^2$, but has a discontinuity for $s>-\Lambda_V^2$. Thus, as remarked in \cite{Caprini:1999ma}, the Landau pole at  $s=-\Lambda_V^2$, present in the  perturbative expansion  (\ref{eq:pertD}) at finite orders, is converted by Borel resummation into an unphysical cut in the region  $-\Lambda_V^2<s<0$, with a discontinuity 
 \beq
 D(s+i\epsilon)- D(s-i\epsilon) = 2\, i \,\mbox{Im}\,D(s+i\epsilon), \quad -\Lambda_V^2<s<0,
 \eeq
 where the imaginary part is given by (\ref{eq:imagdsp}).

\section{Resurgence of the Adler function}\label{sec:resur}
The Borel-\'Ecalle summation of the Adler function, based on the resurgence framework developed in \cite{Costin1995, Costin_Duke, Costin_Book}, was considered in \cite{Maiezza:2021mry} in the large-$\beta_0$ approximation of QCD. As shown in the previous section, in  this approximation the perturbative coefficients are known to all orders and the expression of the Borel transform  $\widehat B_D(u)$ is also known,   its only  singularities being poles on the real axis of the Borel plane. These are precisely the conditions for resurgence required in  \cite{Costin1995, Costin_Duke, Costin_Book}. According to the procedure developed in these works,  the most general solution of the renormalization-group equation obeyed by the Adler function is obtained by supplementing the Borel-summed perturbation series by additional terms, calculated by an iterative algorithm.
 In \cite{Maiezza:2021mry}, the authors applied this algorithm to the QCD Adler function, and derived a resurgent expression of this function   on the spacelike axis $s<0$. In what follows, we shall derive a more general result, writing down the resurgent expression of the  Adler function $D(s)$ valid in the whole complex $s$ plane.

We start from the remark that in our case the algorithm described in \cite{Costin1995, Costin_Duke, Costin_Book} amounts to taking the principal value of the Laplace-Borel summation of the perturbative series and supplementing it by an additional term, given by  the Laplace-Borel integral of the discontinuity  of the Borel transform across the positive axis (the Stokes line) in the Borel plane\footnote{For a recent discussion of the role of the discontinuity of the Borel transform across the Stokes lines in generating transseries to an original divergent series, see also Chapter 2 of \cite{Reis:2022tni} and references therein.}. We show now that this term is obtained immediately in terms of the inverse Mellin
 transform  $\widehat w_D^{(<)}$ defined above. We first set the variable $\tau$ in Eq. (\ref{eq:wn1}) to
\begin{equation}\label{eq:replace}
   \tau=-\frac{\Lambda_V^2}{s} =
   \exp\left( -\frac{4\pi}{\beta_0\,\alpha_s(-s)} \right),
\end{equation}
where we used the one-loop coupling $\alpha_s(-s)$ in the V scheme defined in (\ref{eq:asrng}). Then   (\ref{eq:wn1}) can be written as 
\begin{equation}\label{eq:wn1mod}
   \widehat w_D^{(<)}(-\Lambda_V^2/s) = \frac{1}{2\pi i} \left(\frac{-s}{\Lambda_V^2}\right) \left[\int_{\cal C_+}\!
   du\, \exp\left(-\frac{4\pi u}{\beta_0\alpha_s(-s)} \right)\,\widehat B_D(u) -
   \int_{\cal C_-}\!du\, \exp\left(-\frac{4\pi u}{\beta_0\alpha_s(-s)} \right)\,\widehat B_D(u)\right].
\end{equation}
By comparing this relation with (\ref{eq:Laplace}), we note that the Borel-Laplace of the discontinuity $\widehat B_D(u+i\epsilon)-\widehat B_D(u-i\epsilon)$  for $u>0$ can be identified with the expression
\beq\label{eq:LBdisc}
\frac{\pi}{\beta_0} \left( -\frac{\Lambda_V^2}{s} \right)
   \widehat w_D^{(<)}(-\Lambda_V^2/s).
   \eeq
   
 As follows from \cite{Costin_Duke, Costin_Book},  in the case of a single Stokes line along the positive axis, the function of interest resurges by adding to the perturbative series, summed by a principal-value regulated Laplace-Borel integral, a term equal to the Laplace-Borel integral of the discontinuity of the Borel transform, multiplied by an arbitrary constant.   Therefore, using the Borel resummation (\ref{eq:duplo}) and the expression (\ref{eq:LBdisc}),  we can write down the most general resurgent expression of the Adler function in  the complex $s$-plane
\begin{equation}\label{eq:Dresurg}
   D^{(\pm)}_{resurg}(s) = 1 + \frac{1}{\beta_0} \int\limits_0^\infty\!
   d\tau\,\frac{\widehat w_D(\tau)}{\ln(-\tau s/\Lambda_V^2)} 
   \mp \frac{i\pi}{\beta_0} \left( -\frac{\Lambda_V^2}{s} \right)
   \widehat w_D^{(<)}(-\Lambda_V^2/s) + \frac{c\,\pi}{\beta_0} \left( -\frac{\Lambda_V^2}{s} \right)
   \widehat w_D^{(<)}(-\Lambda_V^2/s),
\end{equation}
where $c$ is an arbitrary real constant. As in (\ref{eq:duplo}), the superscript ``$\pm$'' in parenthesis refers to the sign of 
  $\text{Im}\,s$. Using arguments similar to those applied to the perturbative part in the previous section, one can show that the functions $D^{(\pm)}_{resurg}(s)$ are analytic in the upper/lower half planes of the complex $s$ plane, and  are related by a reflection property similar to Eq. (\ref{eq:Schwarz}). Therefore, they define a single function,  $D_{resurg}(s)$, which is analytic and satisfies the Schwarz reflection property in the whole complex $s$ plane. On the real axis, for $s<-\Lambda_V^2$, the values of  $D^{(+)}_{resurg}(s)$ and $D^{(-)}_{resurg}(s)$ are real and equal. Thus, for $s<-\Lambda_V^2$, we can write
  \beq\label{eq:balance}
   D_{resurg}(s)=\frac{1}{2}\,\left[ D^{(+)}_{resurg}(s)+D^{(-)}_{resurg}(s)\right].
  \eeq 
On the other hand, for $s>-\Lambda_V^2$ the function   $D_{resurg}(s)$ has a cut, and is singular at $s=0$, as seen from the analytic continuation of $\widehat w_D^{(<)}(\tau)$ for $\tau>1$.

We emphasize that the last term in (\ref{eq:Dresurg}) is a purely nonperturbative contribution\footnote{A comparison of this term with the similar quantity derived in \cite{Maiezza:2021mry, Maiezza:2021bed} is not straightforward, due to some possible misprints in these papers. Thus, a factor of 2 seems to be omitted in the expression of $\beta_0$ quoted below Eq. (3)  of \cite{Maiezza:2021bed}, and Eq. (A7) of the same paper slightly differs from the corresponding Eq. (1.20) from \cite{Costin_Duke}. Moreover, the renormalization scale is taken as $\mu^2=Q^2 e^{-5/3}$ above Eq. (12) of \cite{Maiezza:2021mry}, while in Eq. (11) of the same paper the scale seems to be $\mu^2=Q^2$.}, not visible in perturbation theory. Indeed, from (\ref{eq:replace}) and the expression of $\widehat w_D^{(<)}$ given in (\ref{eq:wDfun}), it follows that this term is singular at the origin $\alpha_s=0$ of the coupling plane,  and all its perturbative coefficients (defined as the derivatives with respect to $\alpha_s$ in the limit $\alpha_s\to 0$) are zero. Alternatively,  the nonperturbative term in (\ref{eq:Dresurg}) is seen to contain power-suppressd terms at large $|s|$, much like  the familiar OPE.  More precisely, using the  behavior 
$\widehat w_D^{(<)}(\tau)=8 \tau +O(\tau^2)$ for $\tau\to 0$ which follows from (\ref{eq:wDfun}), we obtain the asymptotic behavior of the last term in (\ref{eq:Dresurg}) at $Q^2=-s\to\infty$ as
\beq\label{eq:Q4}
8\, \frac{c\,\pi}{\beta_0} \, \frac{\Lambda_V^4}{Q^4}.
\eeq

It is of interest to  compare the present work with the study performed in \cite{Mishima:2016vna}, 
which is also based on a representation of the Adler  function in terms of the inverse Mellin 
transform of the Borel transform in the large-$\beta_0$ approximation.  Crucial in 
Ref. \cite{Mishima:2016vna} is the introduction of the infrared cutoff $\mu_f$ and the 
extraction of an ultraviolet part $D_{UV}$, which is independent of $\mu_f$. 
By comparing Eq. (\ref{eq:Dresurg}) with the representation (41) of \cite{Mishima:2016vna},
 we note that the asymptotic term (\ref{eq:Q4}) might be related to the last term $O(\mu_f^4/Q^4)$ in Eq. (41).
Thus, the arbitrary constant $c$ in the resurgent representation derived in this work can be loosely related to the infrared cutoff $\mu_f$ introduced in \cite{Mishima:2016vna}.

 The resurgent representation  (\ref{eq:Dresurg})  is expected to provide a reasonable physical description of the Adler function at large values of $|s|$. 
 In the next section we shall consider for illustration its application for  the evaluation of the hadronic decay rate of the  $\tau$ lepton.   
\section{Hadronic width of the $\tau$ lepton}\label{sec:tau}
The ratio $R_\tau$ of the total hadronic branching fraction to the electron branching fraction of the $\tau$ lepton is expressed in the Standard Model as 
\be\label{eq:Rtau}
R_\tau= 3 \,S_{\rm EW} (|V_{ud} |^2 + |V_{us}|^2 ) (1 + \delta^{(0)} + \delta'_{\rm EW}+\delta_{m}^{D}),
\ee
where $S_{\rm EW}$ is an electroweak factor, $V_{ud}$ and $V_{us}$ are CKM  matrix elements and the last two terms denote  negligible higher corrections \cite{Beneke:2008ad}.

 The quantity of interest for us is the perturbative QCD correction $\delta^{(0)}$.  As shown in   \cite{Braaten:1988hc, Braaten:1991qm, LeDiberder:1992zhd}, this quantity can be written as a weighted integral of the Adler function along a contour in the complex $s$ plane, taken for convenience to be the circle $|s|=m_\tau^2$. In our normalization, this relation is 
\be\label{eq:delta0}
\delta^{(0)} =  \frac{1}{2\pi i} \oint\limits_{|s|=m_\tau^2}\, \frac{d s}{s}\, \left(1-\frac{s}{m_\tau^2}\right)^3\,\left(1+\frac{s}{m_\tau^2}\right)\, [D(s)-1].
\ee

The $\tau$-hadronic width provides one of the most precise methods of extracting the strong coupling $\alpha_s$. By inserting in (\ref{eq:delta0}) the perturbative expansion of the Adler function and comparing the theoretical calculation of $\delta^{(0)}$  with the experimental value, one can extract the coupling at the scale of $m_\tau$. A large number of works have been devoted to this problem, based on various formulations of perturbation theory (see, for instance, Refs.  \cite{Beneke:2008ad, Davier:2008sk,  Caprini:2011ya, Beneke:2012vb, Pich:2013lsa,  Peris:2016jah, Boito:2016pwf,  Caprini:2020lff}). 

The quantity $\delta^{(0)}$ has been evaluated in the large-$\beta_0$ approximation of perturbative QCD in  \cite{Beneke:2008ad}, where Fig. 3 shows the results obtained with the truncated perturbative expansions up to 14th order,  for $\alpha_s(m_\tau^2)=0.34$ in the $\overline{\text{MS}}$ scheme (the definition of FO and CI perturbation theory is given in Section 3 of the quoted paper). The authors indicate also by a horizontal line the value obtained by inserting in  (\ref{eq:delta0})  the PV-regulated Laplace-Borel integral of the full Borel transform (\ref{eq:BDu}). 

In the present work we shall evaluate $\delta^{(0)}$ using the resurgent expression (\ref{eq:Dresurg}) of the Adler function derived in the previous section, which is valid in the complex $s$ plane. 
Since $m_\tau=1.777\,  \text{GeV}$ is larger than the QCD scale $\Lambda_V$, the integration contour in (\ref{eq:delta0}) is in the range where the expression (\ref{eq:Dresurg}) is meaningful.  We use as in \cite{Beneke:2008ad} the value  $\alpha_s(m_\tau^2)=0.34$ in the $\overline{\text{MS}}$ scheme, from which we obtain $\Lambda_{\overline{\text{MS}}}=0.228\,\text{GeV}$ and, using further the relation (\ref{eq:Lambda}),  $\Lambda_V=0.524\,\text{GeV}$.

We first note that the contribution of the perturbative part in (\ref{eq:Dresurg}) to the integral (\ref{eq:delta0}) is equal to 0.2629, in perfect agreement with the result given in Fig. 3 of \cite{Beneke:2008ad}. This illustrates the renormalization-scheme independence of $\delta^{(0)}$ and confirms in a physical case the validity of the resummed expression of the Adler function in the complex plane, derived in \cite{Caprini:1999ma}. We note further that the contribution to  (\ref{eq:delta0}) of the last term in (\ref{eq:Dresurg}), up to the unknown constant $c$, is equal to 0.0091. Therefore, $\delta^{(0)}$ calculated with the resurgent representation  (\ref{eq:Dresurg}) is written as
\beq\label{eq:delta0res}
\delta^{(0)}_{resurg}=0.2629 + 0.0091 c. 
\eeq
 We quote also the phenomenological value
\be\label{eq:input}
\delta^{(0)}_{\rm phen}=0.1966 \pm 0.0040,
\ee
obtained from the value given in \cite{Bethke:2011tr}, page 25,  and the power-corrections contribution to $\delta^{(0)}$, estimated in \cite{Beneke:2008ad}. Consistency of the two predictions is obtained for
\beq\label{eq:c}
c=-7.26 \pm 0.44.
\eeq 
One may  go further and use   the resurgent expression (\ref{eq:Dresurg}) for the evaluation of other moments of the spectral function relevant for $\tau$ decay, used in many papers for the extraction of the strong coupling $\alpha_s(m_\tau^2)$. One must recall however that the large-$\beta_0$ approximation, while useful to get insights on real QCD, is not precise enough for an accurate extraction of the  strong coupling at the $m_\tau$ scale.

\section{Discussion and conclusions}\label{sec:conc}

The resurgence of the QCD Adler function was investigated recently in \cite{Maiezza:2021mry}, where it was argued that the renormalization-group equation satisfied by this function allows the application of the mathematical resurgence approach from  \cite{Costin1995, Costin_Duke, Costin_Book}.  In the present work, assuming the same framework as in \cite{Maiezza:2021mry}, we generalize the results of this paper, by taking advantage of the compact resummation of the perturbative Adler function in the large-$\beta_0$ limit, derived in \cite {Caprini:1999ma}. Using this summation and resurgence mathematical techniques, we obtain  a compact resurgent representation of the Adler function,  valid in the whole complex momentum plane, not only on the spacelike axis. We bring also several other improvements to the treatment in \cite{Maiezza:2021mry} and the subsequent paper \cite{Maiezza:2021bed}. Thus, our result includes completely the  contribution of the ultraviolet renormalons, omitted in the above works. We also avoid some ad-hoc assumptions adopted in \cite{Maiezza:2021mry, Maiezza:2021bed},  such as setting  the lowest four perturbative coefficients to their exact values in full QCD. Our representation contains a single real arbitrary constant, consistent with the existence of a single Stokes line (the real positive axis in the Borel plane), while the authors of  \cite{Maiezza:2021mry, Maiezza:2021bed} introduced several arbitrary constants which multiply various contributions.  Finally, unlike   Ref. \cite{Maiezza:2021bed}, where a nonperturbative model, finite in the infrared region but  inconsistent with the large-$\beta_0$ approximation, has been adopted for the strong coupling,  we use in a consistent way the one-loop coupling (\ref{eq:asrng}). Clearly, the assumptions mentioned above have been adopted in \cite{Maiezza:2021mry, Maiezza:2021bed} in order to increase the flexibility of the theoretical expression used in phenomenological applications.  However,  they lead to a hybrid model whose theoretical uncertainties are difficult to assess.
 
 By contrast,  our work solves completely the problem of resurgence of the Adler function in the large-$\beta_0$ approximation of QCD, based on the mathematical formalism of \cite{Costin1995,Costin_Duke, Costin_Book},  with no ad-hoc assumptions. Our main result, given in Eq. (\ref{eq:Dresurg}), is  a compact expression of the resurgent Adler function in terms of the inverse Mellin transform $\widehat w_D$ of its Borel function, which depends on  a single unknown real constant $c$. The additional term predicted by resurgence is purely nonperturbative, being singular at the origin $\alpha_s=0$ of the coupling plane. The analytic properties of the resurgent Adler function  in the complex momentum plane are determined by the analytic continuation in the complex plane of the inverse Mellin transform, discussed in Sec. \ref{sec:resum}. The representation (\ref{eq:Dresurg}) satisfies the Schwarz reflection principle and is real on the spacelike axis for $s<-\Lambda_V^2$. The representation has cuts along the timelike axis $s>0$ and the spacelike Landau region $-\Lambda_V^2<s<0$. 
 
One might ask whether the present resurgence formalism can be  applied also to full QCD. The inverse Mellin transform $\widehat w_D$  can be defined in principle  also in this case, but in practice its calculation is not possible, since the expression of the exact Borel transform is not known.  The renormalon-based models which describe the Adler function in terms of a small number of leading singularities in the Borel plane are not accurate enough to justify extension by resurgence. So, for the moment, we consider only the large-$\beta_0$ approximation of perturbative QCD in this formalism.

Is useful to recall that  the large-$\beta_0$ approximation  provides actually useful insights to full QCD, and has been considered as such in various contexts. For instance,  it was  used in \cite{Beneke:2012vb} as a laboratory for
renormalon models,  in \cite{Hoang:2021nlz} for understanding the different  ways of treating the running  coupling in the complex $s$
plane and, quite recently in \cite{Boito:2022fmn}, for the calculation of Higgs-boson decay to two photons. Our work brings a contribution to yet another facet of the large-$\beta_0$  approximation, by supplementing the pure perturbative sum of the Adler function by a nonperturbative term, obtained by formal mathematical techniques. By this, the validity of perturbative QCD was extended to lower energies. As shown in Sec. \ref{sec:tau},  where the resurgent representation of the Adler function was used along the circle $|s|=m_\tau^2$,  reasonable contributions to the $\tau$ hadronic width are predicted in the large-$\beta_0$ approximation both by the pure perturbative sum and the nonperturbative contribution. The investigation of other moments of the spectral function measured in hadronic $\tau$ decay is of interest and will be considered in a future work.

~\vspace{0.cm}
\subsection*{Acknowledgments} I am grateful to Ovidiu Costin for enlightening discussions on the resurgence formalism developed in \cite{Costin1995, Costin_Duke, Costin_Book}.




\begin{thebibliography}{99}

  
\bibitem{Dyson:1952tj}
F.~J. Dyson, Divergence of perturbation theory in quantum
  electrodynamics, \href{https://doi.org/10.1103/PhysRev.85.631}{{Phys.
  Rev. {\bfseries 85}, 631 (1952)}}.
  
\bibitem{tHooft} G. 't Hooft, Can we make sense out of Quantum Chromodynamics?
in \href{https://link.springer.com/book/10.1007/978-1-4684-0991-8}{{\em The Whys of Subnuclear Physics}}, edited by A. Zichichi (Plenum Press, New York, 1979), p. 943-982.

\bibitem{Lautrup:1977hs} 
  B.~E.~Lautrup, On high order estimates in QED, \href{https://www.sciencedirect.com/science/article/abs/pii/0370269377901459?via%3Dihub}
 {Phys.\ Lett.\  {\bf 69B}, 109 (1977)}.
 
\bibitem{Broadhurst:1992si}
D.~J. Broadhurst, Large N expansion of QED: asymptotic photon propagator
  and contributions to the muon anomaly, for any number of loops,
  \href {https://doi.org/10.1007/BF01560355} {Z. Phys. {\bf C58}, 339 (1993)}.


\bibitem{Beneke:1992ch}
M.~Beneke, Large order perturbation theory for a physical quantity,
  \href{https://doi.org/10.1016/0550-3213(93)90554-3}{Nucl. Phys.
  {\bf B405}, 424 (1993)}.
  
\bibitem{Beneke:1994qe}
M.~Beneke and V.~M. Braun, Naive nonabelianization and resummation of
  fermion bubble chains,
  \href{https://doi.org/10.1016/0370-2693(95)00184-M}{Phys. Lett. {\bf B348}, 513 (1995)},
  \href{https://arxiv.org/abs/hep-ph/9411229}{arXiv:hep-ph/9411229}.

 
\bibitem{Beneke:1998ui}
M.~Beneke, Renormalons,
  \href{https://doi.org/10.1016/S0370-1573(98)00130-6}{{Phys. Rept.}
  {\bf 317}, 1 (1999)},  \href{https://arxiv.org/abs/hep-ph/9807443}{arXiv:hep-ph/9807443}.

\bibitem{Ecalle1993}
J.~\'Ecalle,  \href{https://link.springer.com/chapter/10.1007/978-94-015-8238-4_3}{Six Lectures on Transseries, Analysable Functions and the Constructive Proof of Dulac's Conjecture, 1993, pp. 75-184}.

\bibitem{BerryHowls} M.V. Berry and C.J. Howls, Hyperasymptotics, \href{https://royalsocietypublishing.org/doi/abs/10.1098/rspa.1990.0111}{Proceedings of the Royal  Society A: Mathematical, Physical and Engineering Sciences {\bf 439}, 653 (1990)}.

\bibitem{Howls} C.J. Howls, An introduction to hyperasymptotics using  Borel-Laplace transforms, in the series \href{https://repository.kulib.kyoto-u.ac.jp/dspace/handle/2433/60642}{Algebraic Analysis of Singular Perturbations}, Kyoto Univ. (1996). 

\bibitem{Dorigoni:2014hea}  D.~Dorigoni, An introduction to resurgence, transseries and alien calculus, \href{https://www.sciencedirect.com/science/article/abs/pii/S0003491619301691?via%3Dihub}{Annals of Phys.\  {\bf 409}, 167914 (2019)}, \href{https://arxiv.org/abs/1411.3585}{{arXiv:1411.3585}}.

\bibitem{Costin1995}
O.~Costin, Exponential asymptotics, transseries, and generalized Borel summation for analytic, nonlinear, rank-one systems of ordinary differential equations, \href{http://dx.doi.org/10.1155/s1073792895000286}{\emph{International
  Mathematics Research Notices} {\bf 1995}, 377 (1995)}.


\bibitem{Costin_Duke} O. Costin, On Borel summati on and Stokes phenomena for rank-1 nonlinear systems of ordinary
differential equations,  \href{http://dx.doi.org/10.1215/S0012-7094-98-09311-5}{Duke Math. J. 93 (2) (1998)}.


\bibitem{Costin_Book}
O.~Costin,   Asymptotics and Borel summability, \href{https://www.taylorfrancis.com/books/mono/10.1201/9781420070323/asymptotics-borel-summability-ovidiu-costin} {Monographs and Surveys in
  Pure and Applied Mathematics. Chapman and Hall/CRC (2008)}.

\bibitem{Aniceto:2018bis}
I.~Aniceto, G.~Basar and R.~Schiappa, A primer on resurgent transseries
  and their asymptotics,
  \href{http://dx.doi.org/10.1016/j.physrep.2019.02.003}{Phys. Rept.
  {\bf 809}, 1 (2019)}, \href{http://arxiv.org/abs/1802.10441}{arXiv:1802.10441}.
  
\bibitem{Reis:2022tni}
T.~Reis, On the resurgence of renormalons in integrable theories,
\href{https://arxiv.org/abs/2209.15386}{arXiv:2209.15386}.

\bibitem{Argyres:2012ka}
P.~C. Argyres and M.~Unsal, The semi-classical expansion and resurgence
  in gauge theories: new perturbative, instanton, bion, and renormalon
  effects, \href{http://dx.doi.org/10.1007/JHEP08(2012)063}{JHEP {\bf
  08}, 063 (2012)}, \href{http://arxiv.org/abs/1206.1890}{arXiv:1206.1890}.

\bibitem{Dunne:2012ae}
G.~V. Dunne and M.~Unsal, Resurgence and trans-series in Quantum Field
  Theory: the CP$^{N-1}$ model,
  \href{http://dx.doi.org/10.1007/JHEP11(2012)170}{JHEP {\bf 11},  170 (2012)}, 
  \href{http://arxiv.org/abs/1210.2423}{arXiv:1210.2423}.

\bibitem{Clavier:2019sph}
P.~J. Clavier, Borel-\'Ecalle resummation of a two-point function, \href{https://link.springer.com/article/10.1007/s00023-021-01057-w}   {Annales Henri Poincare \textbf{22},  2103 (2021)},
  \href{http://arxiv.org/abs/1912.03237}{arXiv:1912.03237}.

\bibitem{Borinsky:2020vae}
M.~Borinsky and G.~V. Dunne, Non-perturbative completion of
  Hopf-algebraic Dyson-Schwinger equation,
  \href{http://dx.doi.org/10.1016/j.nuclphysb.2020.115096}{Nucl. Phys. B
  {\bf 957} (2020) 115096}, \href{http://arxiv.org/abs/2005.04265}{arXiv:
  2005.04265}.

\bibitem{Abbott:2020qnl}
M.~C.~Abbott, Z.~Bajnok, J.~Balog, \'A.~Heged\'{u}s and S.~Sadeghian,
Resurgence in the O(4) sigma model,
\href{https://link.springer.com/article/10.1007/JHEP05(2021)253}{JHEP \textbf{05}, 253 (2021)},
\href{https://arxiv.org/abs/2011.12254}
{arXiv:2011.12254}.

\bibitem{Fujimori:2021oqg}
T.~Fujimori, M.~Honda, S.~Kamata, T.~Misumi, N.~Sakai and T.~Yoda,
  Quantum phase transition and resurgence: Lessons from three-dimensional
  $\mathcal{N}=4$ supersymmetric QED, \href{https://academic.oup.com/ptep/article/2021/10/103B04/6321242?login=false}{PTEP \textbf{2021}, 103B04 (2021)}, \href{http://arxiv.org/abs/2103.13654}{
  arXiv:2103.13654}.

\bibitem{Borinsky:2022knn}
M.~Borinsky and D.~Broadhurst,
Resonant resurgent asymptotics from quantum field theory,
\href{https://www.sciencedirect.com/science/article/pii/S0550321322002127?via%3Dihub}{Nucl. Phys. B \textbf{981}, 115861 (2022)},  
 \href{https://arxiv.org/abs/2202.01513}
{arXiv:2202.01513}.

\bibitem{Ayala:2019uaw}
C.~Ayala, X.~Lobregat and A.~Pineda, Superasymptotic and hyperasymptotic approximation to the operator product expansion, \href{https://journals.aps.org/prd/abstract/10.1103/PhysRevD.99.074019}{Phys. Rev. D \textbf{99},  074019 (2019)}, \href{https://arxiv.org/abs/1902.07736}{arXiv:1902.07736}.

\bibitem{Ayala:2019hkn}
C. Ayala, X. Lobregat and A. Pineda,
Hyperasymptotic approximation to the top, bottom and charm pole mass,
\href{https://journals.aps.org/prd/abstract/10.1103/PhysRevD.101.034002}{Phys. Rev. D \textbf{101}, 034002 (2020)}
\href{https://arxiv.org/abs/1909.01370}{arXiv:1909.01370}.

\bibitem{Ayala:2020pxq}
C. Ayala, X. Lobregat and A. Pineda, Hyperasymptotic approximation to the plaquette and determination of the gluon condensate,
\href{https://link.springer.com/article/10.1007/JHEP12(2020)093}{JHEP \textbf{12}, 093 (2020)}, \href{https://arxiv.org/abs/2009.01285}{arXiv:2009.01285}.

  \bibitem{Maiezza:2021mry}
A. Maiezza and J. C. Vasquez, Resurgence of the QCD Adler function, \href{https://www.sciencedirect.com/science/article/pii/S0370269321002781?via%3Dihub}{Phys. Lett. B \textbf{817}, 136338 (2021)}, \href{https://arxiv.org/abs/2104.03095}{arXiv:2104.03095}.

\bibitem{Maiezza:2021bed}
A. Maiezza and J. C. Vasquez, The QCD Adler function and the muon $g-2$ anomaly from renormalons,   \href{https://www.mdpi.com/2073-8994/14/9/1878}{Symmetry \textbf{14}, 1878 (2022)}, \href{https://arxiv.org/abs/2111.06792}{arXiv:2111.06792}.

\bibitem{Maiezza:2019dht}
A.~Maiezza and J.~C.~Vasquez, non-local lagrangians from renormalons and analyzable functions,
\href{https://www.sciencedirect.com/science/article/abs/pii/S0003491619301113?via%3Dihub}{Annals Phys. \textbf{407}, 78 (2019)}, 
\href{https://arxiv.org/abs/1902.05847}{arXiv:1902.05847}.

\bibitem{Bersini:2019axn}
J.~Bersini, A.~Maiezza and J.~C.~Vasquez,
Resurgence of the renormalization group equation,
\href{https://www.sciencedirect.com/science/article/abs/pii/S0003491620300592?via%3Dihub}{Annals Phys. \textbf{415}, 168126 (2020)},
\href{https://arxiv.org/abs/1910.14507}{arXiv:1910.14507}.

\bibitem{Shifman:1978bx} 
  M.~A.~Shifman, A.~I.~Vainshtein and V.~I.~Zakharov, QCD and resonance physics, 
\href{https://www.sciencedirect.com/science/article/pii/0550321379900221?via%3Dihub}{Nucl.\ Phys.\ B {\bf 147}, 385 (1979)},
 \href{https://www.sciencedirect.com/science/article/pii/0550321379900233?via%3Dihub}{{\bf 147}, 448 (1979)}.

\bibitem{Shifman:2000jv}
M.~A. Shifman, Quark-hadron duality,  in  \href{https://doi.org/10.1142/9789812810458_0032}{At the frontier of particle physics},  pp.~1447--1494, (World Scientific, Singapore, 2001),
  \href{https://arxiv.org/abs/hep-ph/0009131}{arXiv:hep-ph/0009131}.


\bibitem{Boito:2017cnp}
D.~Boito, I.~Caprini, M.~Golterman, K.~Maltman and S.~Peris,
  {{Hyperasymptotics and quark-hadron duality violations in QCD}},
  \href{https://doi.org/10.1103/PhysRevD.97.054007}{{Phys. Rev. D}
  {\bf 97}, 054007 (2018)}, \href{https://arxiv.org/abs/1711.10316}{arXiv:1711.10316}.
  
\bibitem{Caprini:1998wg}
  I. Caprini and J. Fischer, Accelerated convergence of perturbative QCD by optimal conformal mapping of the Borel plane, \href{https://journals.aps.org/prd/abstract/10.1103/PhysRevD.60.054014} {{Phys.\ Rev.}\  D {\bf 60}, 054014 (1999)}, \href{https://arxiv.org/abs/hep-ph/9811367}{arXiv:hep-ph/9811367}.

\bibitem{Caprini:2011ya}
I.~Caprini and J.~Fischer, Expansion functions in perturbative QCD and
  the determination of $\alpha_s(M_\tau^2)$,
  \href{https://doi.org/10.1103/PhysRevD.84.054019}{{Phys. Rev.}
  {\bf D84},  054019 (2011)}, \href{https://arxiv.org/abs/1106.5336}{arXiv:1106.5336}.

\bibitem{Caprini:2020lff}
I.~Caprini, Conformal mapping of the Borel plane: Going beyond perturbative QCD,
\href{https://journals.aps.org/prd/abstract/10.1103/PhysRevD.102.054017}{Phys. Rev. D {\bf 102},  054017 (2020)},
\href{https://arxiv.org/abs/2006.16605}{arXiv:2006.16605}.

\bibitem{Caprini:1999ma} 
  I.~Caprini and M.~Neubert, Borel summation and momentum plane analyticity in perturbative QCD, 
\href{https://iopscience.iop.org/article/10.1088/1126-6708/1999/03/007}{JHEP {\bf 03}, 007 (1999)}, 
 \href{https://arxiv.org/abs/hep-ph/9902244}{arXiv:hep-ph/9902244}.

\bibitem{Baikov:2008jh}
P.~A. Baikov, K.~G. Chetyrkin and J.~H. K\"uhn, Order $\alpha^4_s$ QCD
  corrections to $Z$ and $\tau$ decays, \href{https://doi.org/10.1103/PhysRevLett.101.012002}{Phys. Rev. Lett.
  {\bf 101},  012002  (2008)}, \href{https://arxiv.org/abs/0801.1821}{arXiv:0801.1821}.

\bibitem{Fischer:1994mm}
J.~Fischer, Large order estimates in perturbative QCD and nonBorel summable series,
Fortsch. Phys. \textbf{42}, 665 (1994).

\bibitem{Mueller1985} A. H. Mueller, On the structure of infrared renormalons in physical processes at high energies, \href{https://www.sciencedirect.com/science/article/pii/0550321385904857?via%3Dihub}{Nucl. Phys. {\bf B250}, 327 (1985)}.

\bibitem{Beneke:1997qd}
M.~Beneke, V.~M.~Braun and N.~Kivel, Large order behavior due to ultraviolet renormalons in QCD,
\href{https://www.sciencedirect.com/science/article/abs/pii/S0370269397005625?via%3Dihub}{Phys. Lett. B \textbf{404}, 315 (1997)}, \href{https://arxiv.org/abs/hep-ph/9703389}
{arXiv:hep-ph/9703389}.


\bibitem{Beneke:2008ad}
M.~Beneke and M.~Jamin, $\alpha_s$ and the $\tau$ hadronic width:
  fixed-order, contour-improved and higher-order perturbation theory,
  \href{https://doi.org/10.1088/1126-6708/2008/09/044}{{JHEP} {\bf
  09}, 044  (2008)}, \href{https://arxiv.org/abs/0806.3156}{arXiv:0806.3156}.


\bibitem{Beneke:2012vb}
M.~Beneke, D.~Boito and M.~Jamin,  Perturbative expansion of $\tau$ hadronic
  spectral function moments and $\alpha_s$ extractions,
  \href{https://doi.org/10.1007/JHEP01(2013)125}{{JHEP} {\bf 01},
  125  (2013)}, \href{https://arxiv.org/abs/1210.8038}{arXiv:1210.8038}.


\bibitem{Cvetic:2018qxs}
G.~Cveti\v{c},
Renormalon-motivated evaluation of QCD observables,
\href{https://journals.aps.org/prd/abstract/10.1103/PhysRevD.99.014028}{Phys. Rev. D \textbf{99}, 014028 (2019)},
\href{https://arxiv.org/abs/1812.01580}{arXiv:1812.01}.

\bibitem{Neubert:1994vb}
M.~Neubert, Scale setting in QCD and the momentum flow in Feynman diagrams,
\href{https://journals.aps.org/prd/abstract/10.1103/PhysRevD.51.5924}{Phys. Rev. D \textbf{51}, 5924 (1995)}, 
\href{https://arxiv.org/abs/hep-ph/9412265https://arxiv.org/abs/hep-ph/9412265}{arXiv:hep-ph/9412265}.

\bibitem{Mishima:2016vna}
G. Mishima, Y. Sumino and H. Takaura,
Subtracting infrared renormalons from Wilson coefficients: Uniqueness and power dependences on $\Lambda_{QCD}$,
\href{https://journals.aps.org/prd/abstract/10.1103/PhysRevD.95.114016}{Phys. Rev. D \textbf{95}, 114016 (2017)},
\href{https://arxiv.org/abs/1612.08711}
{arXiv:1612.08711}.

\bibitem{Braaten:1988hc}
E.~Braaten, QCD predictions for the decay of the $\tau$ lepton, \href{https://doi.org/10.1103/PhysRevLett.60.1606}{Phys. Rev. Lett.
  {\bf 60},  1606 (1988)}.

\bibitem{Braaten:1991qm}
E.~Braaten, S.~Narison, and A.~Pich, QCD analysis of the $\tau$ hadronic
  width, \href{https://doi.org/10.1016/0550-3213(92)90267-F}{Nucl.
  Phys. {\bf B373}, 581  (1992)}.
  
\bibitem{LeDiberder:1992zhd}
F.~Le~Diberder and A.~Pich, Testing QCD with $\tau$ decays,
  \href{https://doi.org/10.1016/0370-2693(92)91380-R}{Phys. Lett.
  {\bfseries B289}, 165 (1992)}.

\bibitem{Davier:2008sk}
M.~Davier, S.~Descotes-Genon, A.~Hocker, B.~Malaescu and Z.~Zhang, The determination of $\alpha_s$ from $\tau$ decays revisited, \href{https://link.springer.com/article/10.1140/epjc/s10052-008-0666-7}{Eur. Phys. J. C \textbf{56}, 305 (2008)}, \href{https://arxiv.org/abs/0803.0979}{arXiv:0803.0979}. 

\bibitem{Pich:2013lsa}
A.~Pich, Precision tau physics, \href{https://www.sciencedirect.com/science/article/abs/pii/S0146641013001087?via%3Dihub}{Prog. Part. Nucl. Phys. \textbf{75}, 41 (2014)}, \href{https://arxiv.org/abs/1310.7922}{arXiv:1310.7922}.

\bibitem{Peris:2016jah}
S.~Peris, D.~Boito, M.~Golterman and K.~Maltman, {{The case for duality
  violations in the analysis of hadronic $\tau$ decays}},
  \href{https://doi.org/10.1142/S0217732316300317}{Mod. Phys. Lett A.
  {\bf 31},  1630031 (2016)}, \href{https://arxiv.org/abs/1606.08898}{arXiv:1606.08898}.


\bibitem{Boito:2016pwf}
D.~Boito, M.~Jamin and R.~Miravitllas, Scheme variations of the QCD
  coupling and hadronic $\tau$ decays,
  \href{https://doi.org/10.1103/PhysRevLett.117.152001}{Phys. Rev. Lett.
  {\bfseries 117}, 152001  (2016)}, \href{https://arxiv.org/abs/1606.06175}{arXiv:1606.06175}.
  
\bibitem{Bethke:2011tr}
S.~Bethke et al, Workshop on precision measurements of alphas,
\href{https://arxiv.org/abs/1110.0016}{arXiv:1110.0016}.


  \bibitem{Hoang:2021nlz}
A.~H.~Hoang and C.~Regner,
On the difference between FOPT and CIPT for hadronic tau decays,
\href{https://link.springer.com/article/10.1140/epjs/s11734-021-00257-z}{Eur. Phys. J. ST \textbf{230},  2625 (2021)},
\href{https://arxiv.org/abs/2105.11222}
{arXiv:2105.11222}.

  \bibitem{Boito:2022fmn}
D.~Boito, G.~das Neves and J.~Piclum,
$H \to \gamma \gamma$ to all orders in $\alpha_S$ in the large-$\beta_0$ limit of QCD,
\href{https://journals.aps.org/prd/abstract/10.1103/PhysRevD.106.094026}{Phys. Rev. D \textbf{106}, 094026 (2022)},  \href{https://arxiv.org/abs/2209.00369}{arXiv:2209.00369}.



\end{thebibliography}
\end{document}